\documentclass[10pt]{article}
\usepackage{a4}

\usepackage{epsfig}

\begin{document}

\title{Solution of multi-channel  problems using MCAS
for spectra and  scattering cross sections.}

\author{K. Amos~$^1$, S. Karataglidis~$^1$, 
P. Fraser~$^1$,
D. van der Knijff~$^2$,\\
J. P. Svenne~$^3$, 
L. Canton~$^4$, and G. Pisent~$^4$}

\maketitle

\begin{abstract}
           A multi-channel algebraic scattering theory, to find solutions of 
coupled-channel  scattering  problems  with  interactions  determined     by 
collective models, has been structured to ensure that the Pauli principle is 
not violated. Positive (scattering) and negative (sub-threshold)   solutions 
can be found to predict both the compound nucleus sub-threshold spectrum and 
all resonances due to coupled-channel effects that occur on a smooth  energy 
varying background.   The role of the Pauli principle is crucial in defining 
what interaction potentials are required to fit data.  The theory also gives 
an algebraic form for the  dynamic  polarization potential which adds to the 
ground state interaction to define the optical potential that gives the same 
elastic scattering cross sections.
\end{abstract}

\footnotetext[1]
{School of Physics, The University of Melbourne, Victoria 3010, Australia}
\footnotetext[2] 
{Advanced  Research   Computing, University of Melbourne, Victoria 3010, 
Australia}
\footnotetext[3] 
{Department  of  Physics  and Astronomy, University  of Manitoba, and\\
\indent\indent
Winnipeg Institute  for   Theoretical  Physics, Winnipeg, Manitoba, Canada 
R3T 2N2}
\footnotetext[4] 
{Istituto  Nazionale  di  Fisica  Nucleare, Sezione  di Padova, e 
Dipartimento di Fisica\\
\indent\indent  dell'Universit\`a
    di Padova, via Marzolo 8, Padova I-35131, Italia}


\section{Introduction}
 
     Low-energy cross sections from the collision of nucleons with light-mass 
nuclei show sharp as well as broad resonances upon a smooth, energy-dependent 
background. Those resonances may correlate to states in the discrete spectrum 
of the target.    To interpret such scattering data requires use of a complex 
coupled-channel reaction theory .
A theory with very important improvements over those used heretofore has been
developed~\cite{Am03}.     This
theory, named a multi-channel algebraic scattering (MCAS) theory, finds solution of 
the coupled Lippmann-Schwinger equations for the quantum    scattering of two 
fragments.  Though the method can be applied to quite general scattering systems, 
to date studies
have been limited to nucleon scattering from zero ground state spin targets.

The prime information sought are  scattering ($S$-) matrices which are easily
extracted from the $T$-matrices generated by MCAS.      The approach involves 
using matrix algebra on matrices built using Sturmian-state expansions of the 
relevant nucleon-nucleus potential matrix.   With this method, all resonances 
in any energy range  can be identified,      and their centroids, widths, and 
spin-parities    determined    as    can    the energies and spin-parities of 
sub-threshold states of the system.

                It has long been known that collective-model prescriptions of 
nucleon-nucleus scattering violate the Pauli principle.     However, the MCAS
procedure    enables    use    of   an orthogonalizing pseudo-potential (OPP) 
approximation by which unphysical effects due to   violation   of  the  Pauli 
principle can be avoided. Doing so is crucial to finding the parameter values
of the interaction that  simultaneously  gives  the  sub-threshold   spectrum
and the low energy scattering cross sections~\cite{Ca05}.

       The multichannel formalism and the process by  which resonances can be 
identified and located are outlined next.  Results of calculations made using 
a collective-model prescription for the potential matrix are then 
discussed. In that collective model, the interaction field is allowed  to  be 
deformed from sphericity and  that deformation is taken to second order.   In
studies of nucleon scattering from $^{12}$C, coupling is taken to the  ground
$0^+_1$, $2^+_1$ (4.4389 MeV), and $0^+_2$ (7.6542 MeV) states  and we consider
nucleon energies to 6 MeV.     With these three states defining the relevant
target spectrum, only quadrupole deformation is involved.         In a new
development, we have applied the MCAS method to the $p$+${}^6$He system     and
have compared results with the spectrum of ${}^7$Li known from experiment.

The theory also allows the construction of the optical potential by 
appropriately
subsuming  the  coupled-channel  equations  into  effective elastic-channel 
scattering equations. The optical potentials thus constructed,     as well as
allowing for the Pauli principle, are very nonlocal and energy    dependent.
However,   the   potentials    guarantee that their use lead to cross section 
details including resonance effects.  We present typical
results for the $n$+${}^{12}$C case.


\section{The MCAS theory (in brief)}
\label{multiT}

The integral equation approach to scattering in momentum space  for potential
matrices $V_{cc'}^{J^\pi}(p,q)$ where $c=[(\ell s) j, I J^\pi]$ where $J^\pi$
are  conserved  quantum  numbers which will often be omitted for convenience,  
requires   solution   of  coupled Lippmann-Schwinger  (LS) equations giving a  
(partial wave)   multichannel $T$-matrix of the form 
\begin{eqnarray}
T_{cc'}(p,q;E)\, = \, V_{cc'}(p,q) +
\frac{2\mu}{\hbar^2}
\left[ \sum_{c'' = 1}^{\rm open} \int_0^\infty V_{cc''}(p,x) 
\frac{x^2}{k^2_{c''} - x^2 + i\epsilon} T_{c''c'}(x,q;E)
\ dx \right.&&
\nonumber\\
\left.- \sum_{c'' = 1}^{\rm closed} \int_0^\infty 
V_{cc''}(p,x) \frac{x^2}{h^2_{c''} + x^2} 
T_{c''c'}(x,q;E) \ dx \right]\ .\:&&
\label{multiTeq}
\end{eqnarray}
The    open    and    closed channels contributions have channel wave numbers 
$k_c = \sqrt{\frac{2\mu}{\hbar^2}(E - \epsilon_c)}$                       and 
$h_c = \sqrt{\frac{2\mu}{\hbar^2}(\epsilon_c - E)}$, for $E > \epsilon_c$ and 
$E < \epsilon_c$ respectively.        $\epsilon_c$ is the threshold energy of
the target state inherent in   channel  $c$  and  $\mu$  is the reduced mass.
Solutions of Eq.~(\ref{multiTeq}) are sought using expansions of chosen model
potential matrix elements in (finite) sums of energy-independent    separable
terms,
\begin{equation}
V_{cc'}(p,q) \sim  \sum^N_{n = 1} \chi_{cn}(p)\ 
\eta^{-1}_n\ \chi_{c'n}(q)\ .
\label{finiteS}
\end{equation}

 In the MCAS method~\cite{Am03}, the form factors are specified in terms of a
set ($n$) of Sturmians $\Phi_{c,n}(p)$ via
\begin{equation}
\left|\chi_{c,n}\right\rangle  
= \sum_{c'} V_{cc'}\left|\Phi_{c,n}\right\rangle\ ,  
\end{equation}
and the Sturmians themselves are solutions of coupled-channel (Schr\"odinger)
equations for the chosen matrix of potentials $V_{cc'}(p,q)$ with eigenvalues
$\eta_n$.

 In coordinate space those potentials are defined using a collective model 
prescription for the nucleon-target system,    with local forms $V_{cc'}(r)$.
The Pauli principle can be satisfied by using an  OPP method in the determination 
of the Sturmians.         Those Sturmians then are solutions  of coupled-channel 
equations for the matrix of nonlocal potentials 
\begin{equation}
\mathcal{V}_{cc'}(r,r')  =\,  V_{cc'}(r)\delta(r-r')\,  +\,  \lambda
    A_c(r) A_c(r')\, \delta_{cc'}\ ,
\end{equation}
where   $A(r)$  is  the radial part of the single particle (bound) state wave 
function  in channel $c$   spanning  the  phase  space  excluded by the Pauli 
principle. The OPP  method holds in the limit $\lambda \to \infty$,   but use 
of $\lambda = 1000$~MeV suffices.

                   The link between the multichannel $T$- and $S$-matrices is
$S_{cc'} = \delta_{cc'} -i \pi \frac{2\mu}{\hbar^2} \sqrt{k_c k_{c'}}\ 
T_{cc'}$  which expands in matrix form with the Sturmians as the basis set to
\begin{equation}
S_{cc'} = \delta_{cc'} - i^{l_{c'} - l_c +1} 
\frac{2\mu\pi}{\hbar^2} \sum_{n,n' = 1}^N 
\sqrt{k_c} \chi_{cn}(k_c) \left([\mbox{\boldmath $\eta$} 
- {\bf G}_0]^{-1} \right)_{nn'} \ \chi_{c'n'}(k_{c'})\sqrt{k_{c'}}\ ,
\label{multiS}
\end{equation}
where now $c,c'$ refer to open channels only.         In this representation, 
\mbox{\boldmath $\eta$}  has  matrix    elements $\left( \eta \right)_{nn'} = 
\eta_n\ \delta_{nn'}$ while \textbf{${\bf G}_0$} has elements 
\begin{eqnarray}
\left( {G}_0 \right)_{nn'} &=& 
\frac{2\mu}{\hbar^2}\left[ \sum_{c = 1}^{\rm open} 
\int_0^\infty \chi_{cn}(x) \frac{x^2}{k_c^2 - x^2 + i\epsilon} 
\chi_{cn'}(x)\ dx  \right.
\nonumber\\
&&\hspace*{1.5cm}\left.
 - \sum_{c = 1}^{\rm closed} \int_0^\infty 
\chi_{cn}(x) \frac{x^2}{h_c^2 + x^2} \chi_{cn'}(x)\ dx 
\right]\ . 
\label{xiGels}
\end{eqnarray}
The   bound   states   of the compound system are defined by the zeros of the 
matrix determinant when the energy is     $E < 0$ and so link to the zeros of 
$\{ \left| \mbox{\boldmath $\eta$}-{\bf G}_0\right| \}$  when all channels in
Eq.~(\ref{xiGels}) are closed.

The Sturmian expansions and their use have been detailed recently~\cite{Am03}
and are not repeated herein.     Suffice it to note that a finite  set can be 
used to make the expansion of  finite rank.  An
essential feature in such studies,        if very narrow resonances are to be 
observed, is a resonance finding scheme.      Such is also given in detail in
Ref.~\cite{Am03}.  Essentially that requires recasting the elastic scattering
$S$-matrix  (for each $J^\pi$ and with the elastic channel taken as $c=1$) as
\begin{equation}
S_{11} = 1 - i \pi \frac{2\mu}{\hbar^2} \sum_{nn'=1}^M k \ \chi_{1n}(k) 
\frac{1}{\sqrt{\eta_n}} \left[\left({\bf 1} -  
\mbox{\boldmath $\eta$}^{-\frac{1}{2}}
{\bf G}_0\mbox{\boldmath $\eta$}^{-\frac{1}{2}}    
\right)^{-1}\right]_{nn'} \frac{1}{\sqrt{\eta_{n'}}} 
\chi_{1n'}(k)\, .
\end{equation}
Here,    the    elements    of    the    diagonal    (complex)         matrix 
\mbox{\boldmath $\eta$}$^{-\frac{1}{2}}$ are
$\frac{1}{\sqrt{\eta_n}}\delta_{nn'}$.    Then the  complex-symmetric matrix, 
\mbox{\boldmath $\eta$}$^{-\frac{1}{2}}{\bf G}_0$ 
\mbox{\boldmath $\eta$}~$^{-\frac{1}{2}}$               can  be  diagonalized 
and the evolution of its complex eigenvalues    $\zeta_r$    with  respect to 
energy, define  resonance attributes.    Resonant behavior occurs when one of
the complex $\zeta_r$ eigenvalues passes close   to  the point $(1,0)$ in the 
Gauss plane.  At the energy corresponding to that point,  the elastic channel 
$S$-matrix has a pole.

\section{Results using MCAS for $n$ and $p$ + ${}^{12}$C}

   The MCAS approach has many advantages over the more usual coordinate space 
solution methods. Notably with it one can
1) account for the effects of the Pauli principle, even for collective model 
descriptions of the coupling potentials,
2) specify sub-threshold states in the compound nucleus,
3) locate all resonances caused by $V_{cc'}(r)$ and find their centroid energies
and widths in the energy range chosen, no matter how small their widths,
and 4) specify  the  complete $T$- and $S$-matrices off-shell enabling prediction of 
all low energy elastic and inelastic cross sections 
and spin observables.

\subsection{A comparative study}

In this first subsection,  we present the results from a comparative study of 
a coordinate space  program  with MCAS  (with and without taking into account 
the Pauli principle).     We seek to see if, when one performs as exactly the
same evaluation as possible,  the calculations are equivalent.    Also, for a
typical low energy problem,     we ask how the Pauli principle influences the
results.

       To make a ``fair'' comparison, we have used a simplified model for the 
$n$+${}^{12}$C system.      We chose the same three target states to define the 
coupled channels in both   the    coordinate-space   (ECIS97~\cite{Ra94}) and
momentum-space (MCAS) evaluations.  For this we take coupling to be effected
by a simple rotational   model   scheme having only a quadrupole deformation,
$\beta_2 = -0.52$, on the real potential~\cite{Ra94}, 
\begin{equation}
V(r) = -49.92\, f(r)\, +\, \left(\frac{\hbar}{m_\pi c}\right)^2
6\, {\mathbf \sigma \cdot \nabla f(r) \times \frac{1}{i}\nabla};\;\;\; 
f(r) = \left[1 + e^{\left(\frac{r - 2.885}{0.63}\right)} \right]^{-1} .
\label{Equation1}
\end{equation}
The potential units are MeV and all lengths are in fm.
In the MCAS  evaluation, however, the spin-orbit term is reduced to  the 
{\bf l$\cdot$s} form. Doing so has little effect 
on elastic scattering~\cite{Am05}.  Initial runs of ECIS97 for the simplified model
\begin{minipage}[t]{6.0cm}
\vspace*{1mm}
\centering{
\epsfig{figure=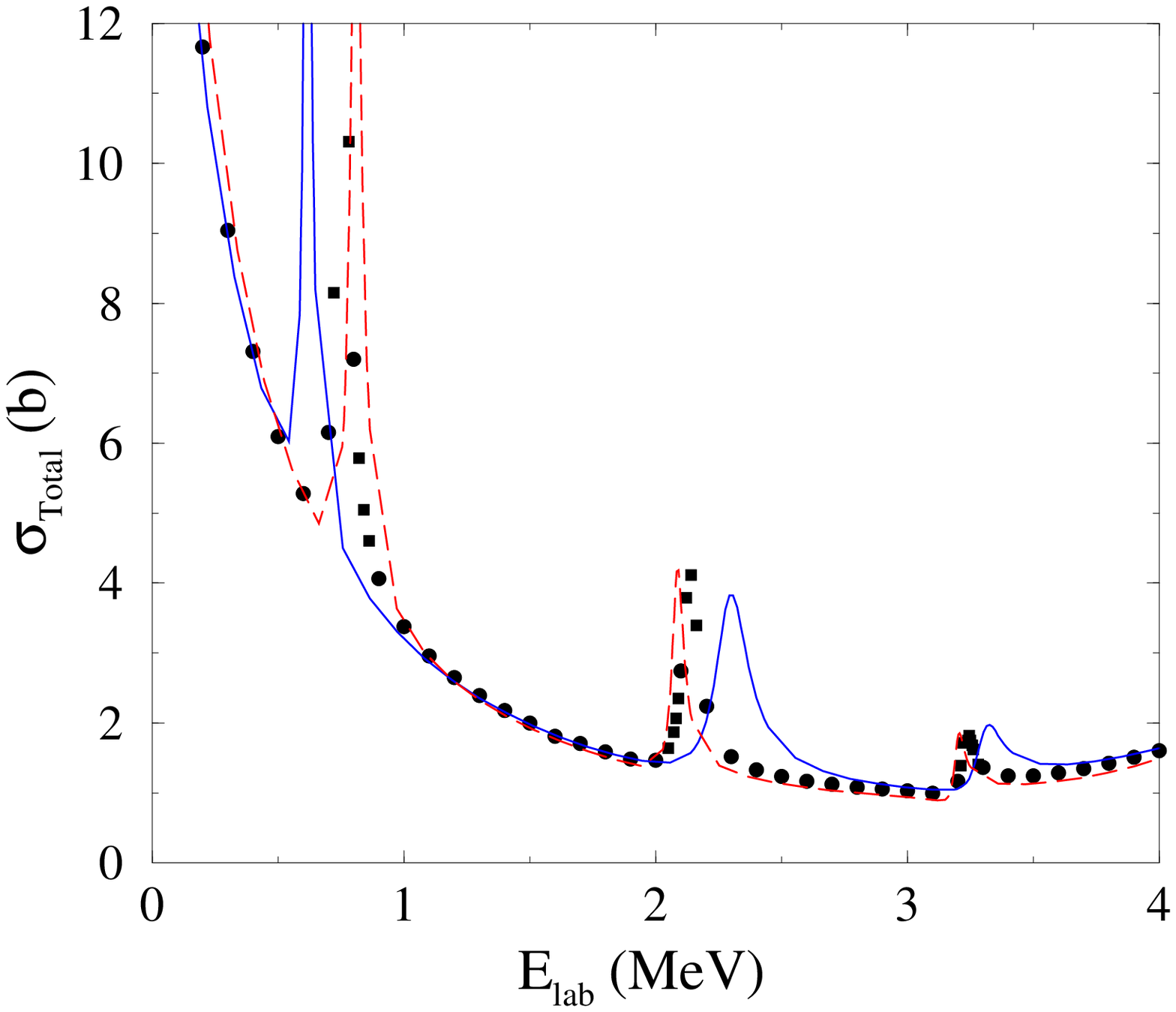,clip=,width=\linewidth}
}
\end{minipage} \hfill
\begin{minipage}[t]{6.3cm}
were for (lab.) energies from 0.1 
MeV to 4.0 MeV in steps of 0.1 MeV.   Those results indicated that there were
three resonances near 0.7, 2.1, and 3.3 MeV.    Thus additional calculations 
were made for more closely spaced energies spanning those three values and
the results are displayed by the filled circles in Fig.~1.\\
 
{\small Figure~1: 
The $n$+${}^{12}$C cross section results from using MCAS theory 
with those found using the ECIS approach. Details are given in the text.}
\end{minipage} \hfill
Therein     they are compared with the results of  MCAS calculations, using the 
same  simplified  model  and,  essentially,  the  fixed  interaction  given  in 
Eq.~(\ref{Equation1}),   but  without  accounting for the Pauli principle.  The
solid curve is the MCAS result when the deformation (of $V_{cc'}$)  was   taken
through second order~\cite{Am03,Pi05}.    The three resonances seen in the ECIS 
results are present though their energy centroids are slightly shifted. But the
background on which those resonances lie agrees very well with the ECIS result.
The second MCAS result (dashed curve) was obtained by limiting deformation   to
first order and now the agreement with the ECIS result is much better for  both
background and the resonances;  centroids and widths.

  We draw two conclusions.   First when the simplified model is used in exactly 
the   same  way  in  finding solutions of the coupled-channel    problem using
the coordinate-space approach~\cite{Ra94} and the momentum-space approach  with
MCAS~\cite{Am03}, the scattering cross sections agree very well.   The   smooth
background as well as the  specific  resonances that can be generated with ECIS
are found with the MCAS run.  Those were calculations with deformation taken to
first order.   The second conclusion from comparison of the two MCAS results is
that, with deformation of -0.52 in the $n$+${}^{12}$C system,        expansion of
$V_{cc'}$ need be taken to second order.

              However, all of the calculations ignored the effects of the Pauli 
principle~\cite{Ca05}.  But in the MCAS approach, the OPP method can be used to
ensure that there is no violation of the Pauli principle even when a collective
model is used to specify the interaction potentials.  The OPP method constrains
the  Sturmians used as an expansion set are orthogonal to all states  in  which
the incoming nucleon would be trapped into an orbit fully occupied by  nucleons
in the target.      Using such a conditioned Sturmian function set to solve the
MCAS theory of coupled equations gave the cross section displayed by the  solid
curve in Fig.~2.      There is a considerable change in the resonances observed 
though the background cross section, being dominated by $s$-wave scattering, is
little varied. \\
\begin{minipage}[t]{5.8cm}
\vspace*{1mm}
\hspace*{-0.6cm}
\epsfig{figure=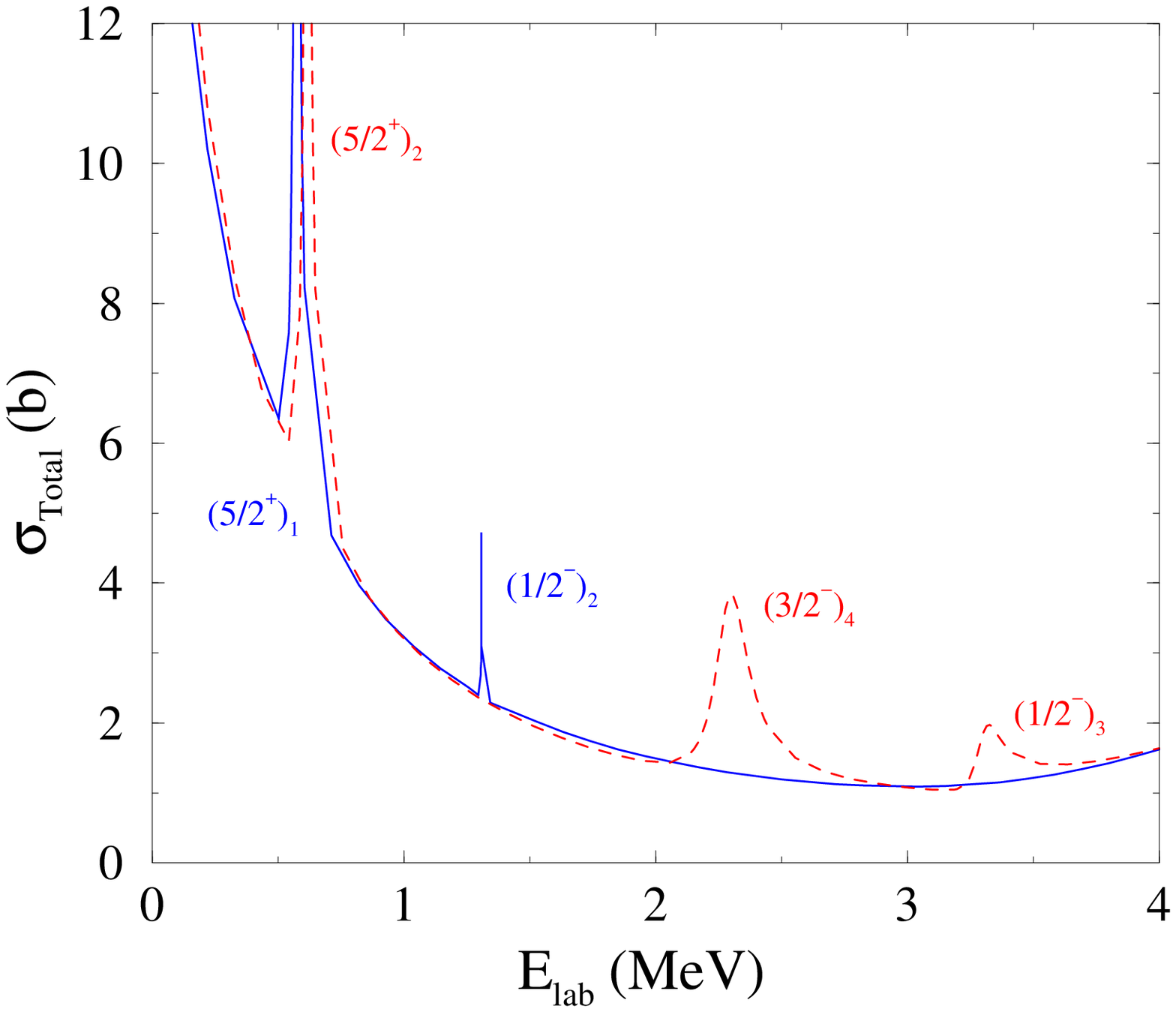,clip=,width=\linewidth}
\vspace*{1mm}
\end{minipage} \hfill
\begin{minipage}[t]{5.8cm}
\vspace*{1mm}
\centering{
\epsfig{figure=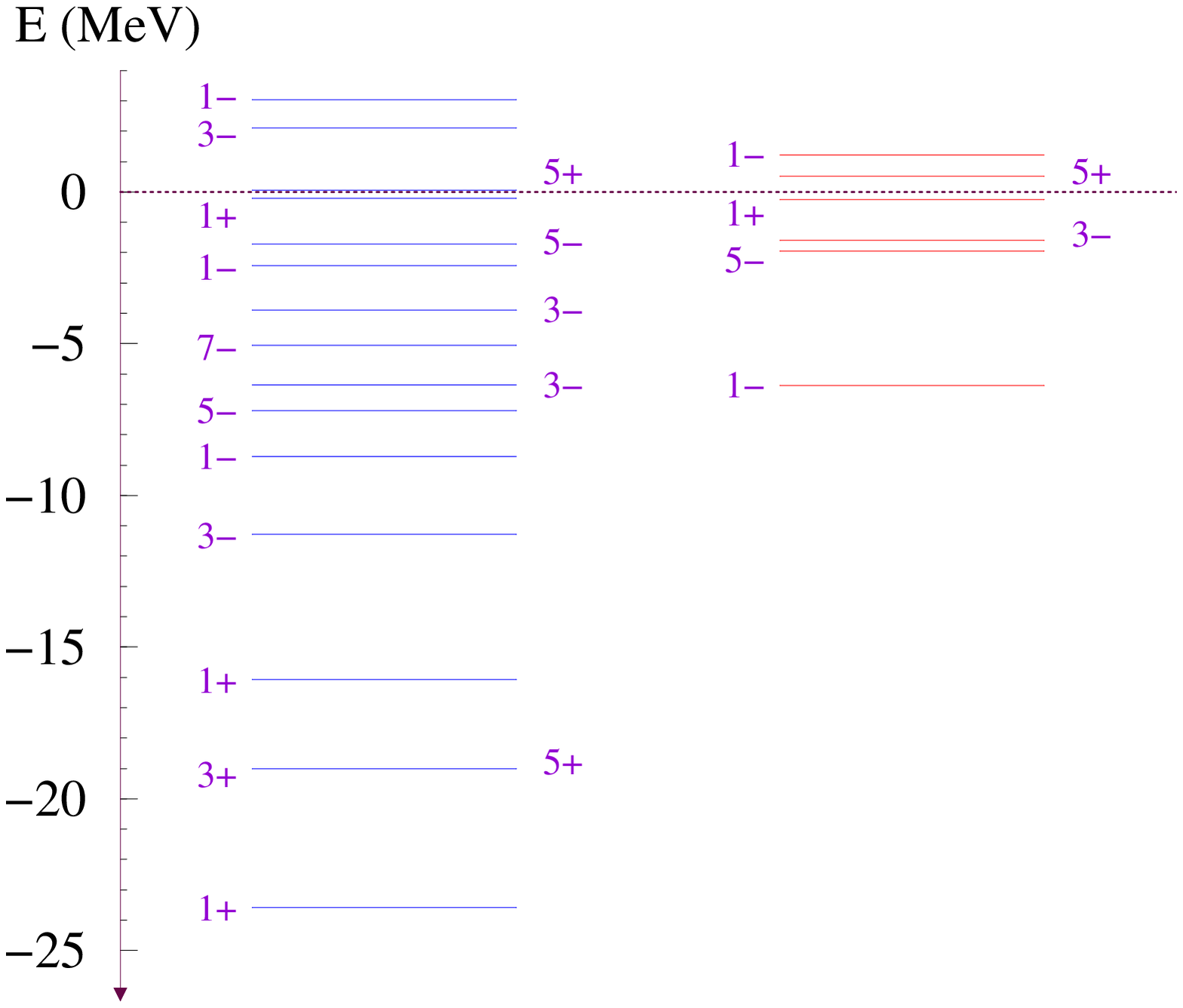,clip=,width=0.98\linewidth}}
\vspace*{1mm}
\end{minipage} \hfill
\begin{minipage}[t]{5.8cm}
{\small Figure~2:
The $n$+${}^{12}$C cross section results from using MCAS theory with (solid 
curve) and without (dashed curve) using the OPP method.} 

\vspace*{1mm}
\end{minipage} \hfill
\begin{minipage}[t]{5.8cm}
{\small  Figure~3:
 The $n$+${}^{12}$C sub-threshold (and resonance) spectra found using MCAS 
with (right) and without (left) accounting for the Pauli
principle.}\\

\vspace*{1mm}
\end{minipage} \hfill

The number of resonances drops from three  to  two  in  this  energy range when
the Pauli principle is considered.   Further,  those resonances inherently have
different underlying structures, as the $\frac{5}{2}^+$ resonance with centroid
near 0.6 MeV is of rank 1 and not 2 when the Pauli principle is     considered.
(Here by rank we mean the number of the state of  given $J^\pi$ in ${}^{13}$C.)
Likewise the $\frac{1}{2}^-$ resonance,         in addition to being shifted in 
position by some 2 MeV,  has a smaller width and is of rank 2.

       The  MCAS method can be used with negative energies and   so produce the 
sub-threshold compound nucleus bound state spectrum. With the fixed interaction
of Eq.~(\ref{Equation1}), the spectra    (including the resonance states) found
with    and   without using the OPP are shown on the  right and  left of Fig.~3 
respectively.   If the Pauli principle is violated,  a large number of spurious 
states result. 

\subsection{Resonance finding}
  With MCAS, all resonances (spin-parity, centroid, width) can be found without 
requiring use of an inordinately large number of energy values.   The procedure
is   to   solve  the  matrix equations for the eigenvalues $\zeta_r(E_i)$ 
for a (finite) set of 
energies.         The energy centroid of a resonance is that value at which the 
eigenvalue track passes close to the point (1,0).  Those eigenvalues are complex and 
their imaginary part relates to the resonance width~\cite{Am03}.   A sample set
of  the  eigenvalues  for $\frac{3}{2}^+$ resonances in the $n$+${}^{12}$C system 
generated using this method are shown in Fig.~4.        The point values of the
largest eigenvalue (labeled (1)) has been connected by a long-dashed line    to
guide the eye to link the energy sequence of the results. The actual trajectory
has a cusp. That is very evident with curve (1), but similar though more slight
features are evident in the trajectories (2) and (3).  That cusp feature of the
eigenvalue trajectories links to the opening of the $2^+_1$ state at 4.4389 MeV
in the coupled-channel algebra.          Note that these eigenvalues have small
imaginary parts and so the plot is semi-logarithmic.   Thus some details of the
trajectories lie below the graphed range.        The largest eigenvalue clearly
evolves well beyond the unit circle after crossing at the energy of the  lowest
$\frac{3}{2}^+$ compound  resonance in this cross section.     That centroid is
located at 3.0 MeV.     The imaginary part is consistent with the width. $\;\;$
Next note\\
\begin{minipage}[t]{6.0cm}
\vspace*{0.001mm}
\centering{
\epsfig{figure=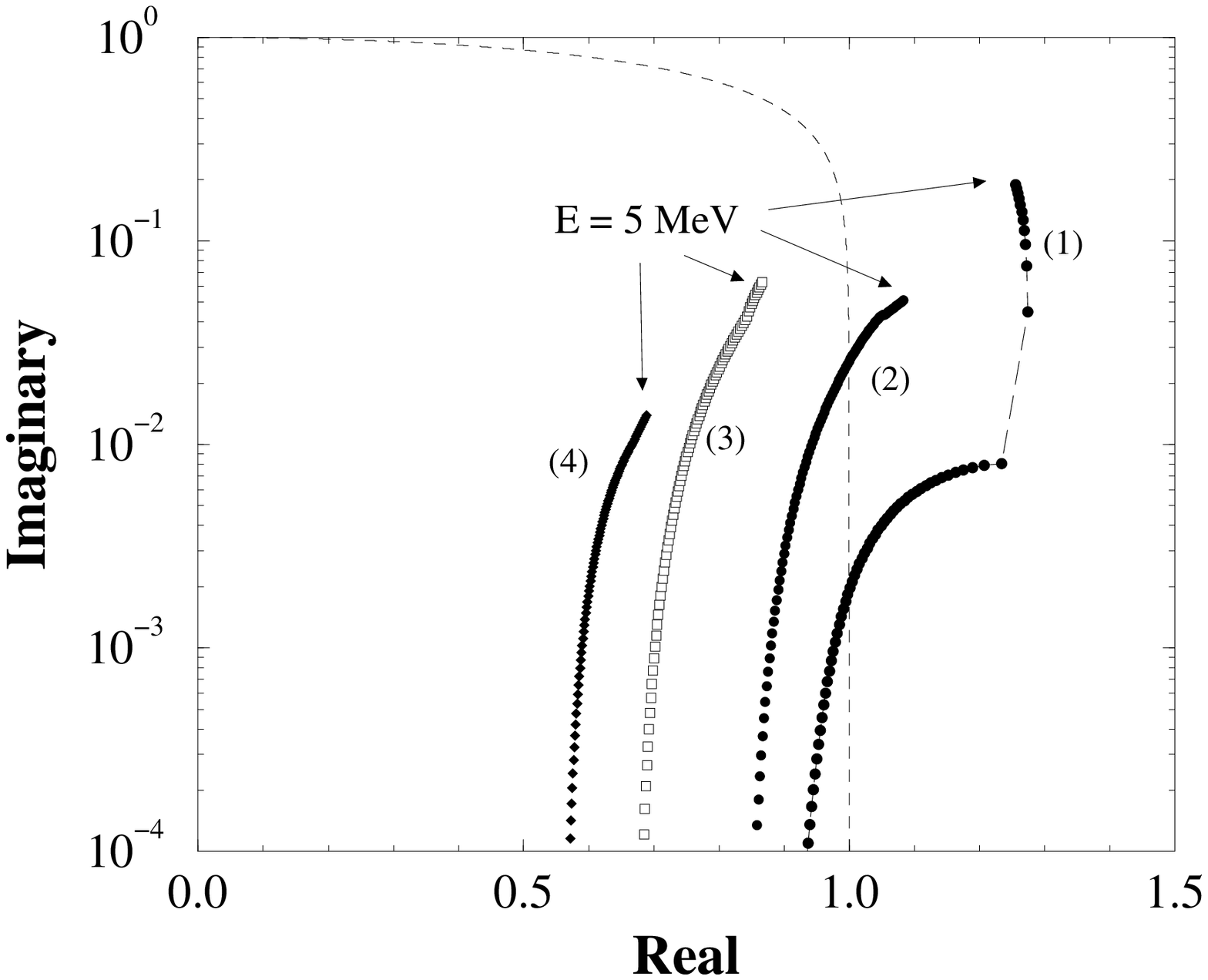,clip=,width=1.05\linewidth}
}
\end{minipage} \hfill
\begin{minipage}[t]{6.0cm}
that the 2$^{\rm nd}$ trajectory crosses the unit circle at higher energy   and
with a larger imaginary part.            This coincides with a second, broader,
$\frac{3}{2}^+$ resonance at 3.4 MeV.         The 3$^{\rm rd}$ and 4$^{\rm th}$
Sturmian  trajectories shown  track toward the unit circle but have not crossed
before 5 MeV.    In this way all resonances of all the spin-parities considered
are always found, no matter how narrow they are.\\

{\small  Figure~4:
Argand plots of $\frac{3}{2}^+$ resonance finding eigenvalues
for the $n$+${}^{12}$C system. The dashed line is the unit circle.}
\end{minipage}\hfill

\subsection{Cross sections and sub-threshold states}

 Results of our MCAS study of the neutron plus ${}^{12}$C system for excitation 
in ${}^{13}$C to $\sim 10$ MeV, are displayed in Fig.~5.        The interaction
potential parameters used were
\begin{minipage}[t]{7.0cm}
\vspace*{0.1mm}

\epsfig{figure=Amos-BIII-fig2.eps,clip=,width=\linewidth}

\vspace*{0.1mm}
\end{minipage} \hfill
\begin{minipage}[t]{5.0cm}

those tabulated in Ref.~\cite{Ca05}.    In this figure,  the elastic scattering
cross section is compared with data with   respect   to   the   ground state of 
${}^{12}$C as zero energy.  On the same scale,   in the right hand parts of the 
figure, we compare the experimental with theoretical sub-threshold  and
\\

\vspace*{1mm}

{\small Figure~5:
Spectra of     ${}^{12,13}$C and the elastic cross section for the $n$+${}^{12}$C 
system.} 

\vspace*{2mm}
\end{minipage}\hfill
resonance states in ${}^{13}$C. $\;$  The labels
denote values of $2J$ and parity.     

     That same elastic cross section and MCAS result are shown in the top panel
of Fig.~6 with the bottom two panels showing what using the   $S$-matrices give
as predictions of polarizations at two scattering angles.         The resonance
structures most evident in the cross section are well reproduced by MCAS    and
those details are confirmed in the polarizations.      The spin-parities of the
resonances   also   coincide   with   those  of known states in the spectrum of 
${}^{13}$N.   Notably the prominent $\frac{5}{2}^+$ resonance at 2 MeV, and the 
two $\frac{3}{2}^+$ resonances spanning 2.8 - 4.0 MeV, are found with very good
widths, narrow and broad respectively.       In Fig.~7, proton scattering cross
sections    (from ${}^{12}$C) is shown at two scattering angles.   Found simply
by adding a Coulomb potential to the $n$+${}^{12}$C system interactions,      the
results from MCAS calculations (of $p$-${}^{12}$C scattering) compare well   with
the data.

\subsection{The spectra of ${}^7$Li from the $p$+${}^6$He system}

        MCAS results for the $p$+${}^6$He system have been found  when the target 
(${}^6$He)  structure  is  taken  as  three states; the $0^+$ (ground), and two 
$2^+$ states at 1.78 and 5.68\\
\begin{minipage}[t]{6.0cm}
\vspace*{1mm}

\epsfig{figure=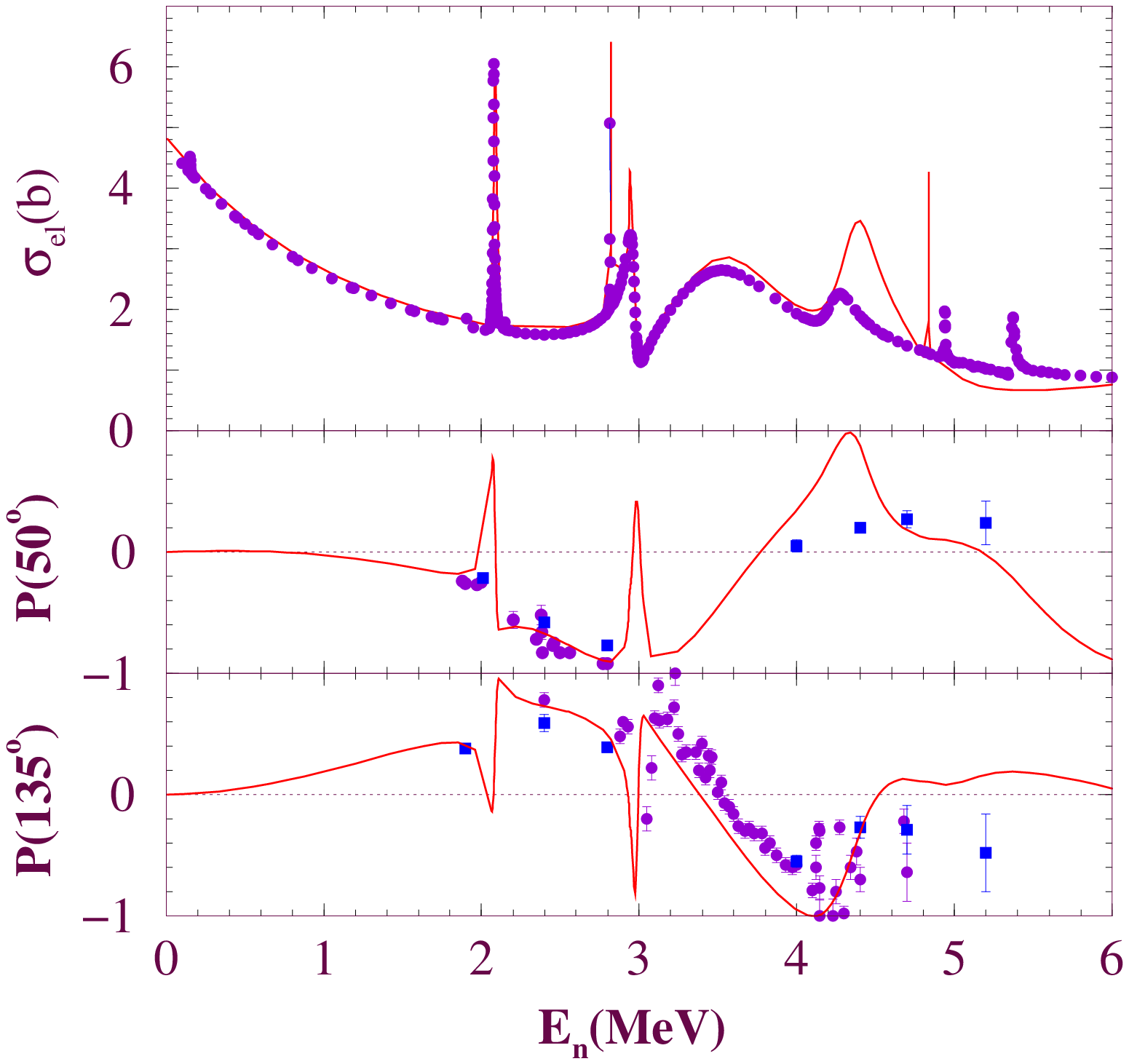,clip=,width=0.9\linewidth}

\vspace*{1mm}
\end{minipage} \hfill
\begin{minipage}[t]{6.0cm}
\vspace*{1mm}

\epsfig{figure=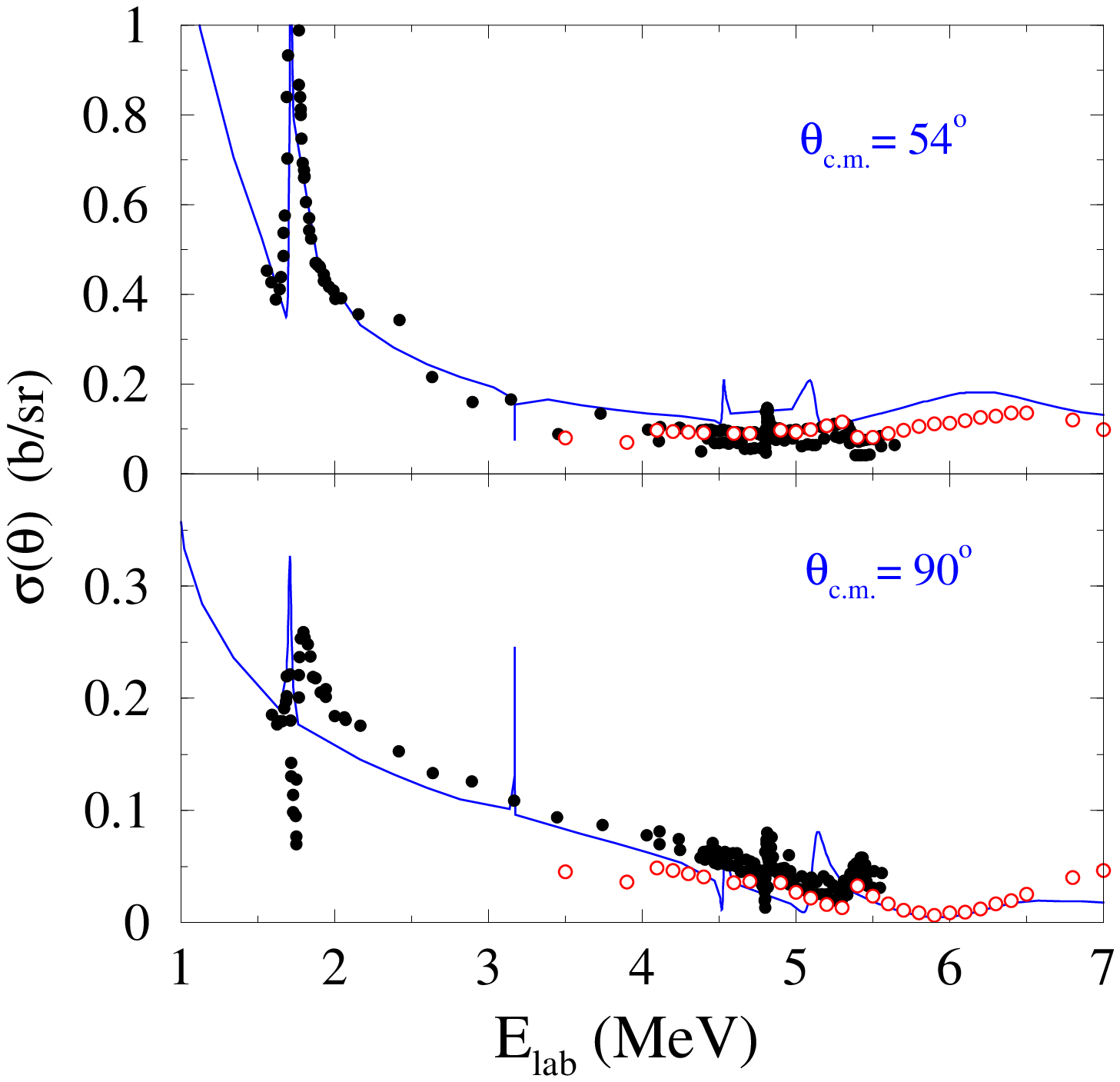,clip=,width=0.9\linewidth}

\vspace*{1mm}
\end{minipage} \hfill
\begin{minipage}[t]{6.0cm}
\vspace*{1mm}

{\small Figure~6:
The   elastic cross section for n-${}^{12}$C scattering (top) and polarizations 
at 50$^\circ$ (middle) and at 135$^\circ$ (bottom).}
\vspace*{2mm}
\end{minipage} \hfill
\begin{minipage}[t]{6.0cm}
\vspace*{1mm}

{\small Figure~7:
MCAS    predictions    and    proton scattering cross-section data at two fixed 
scattering angles.}
\vspace*{2mm}
\end{minipage} \hfill
MeV respectively. A quadrupole deformation was considered and both $2^+$ states
were allowed to couple to the ground with the same deformation and between each
other through second order.   To get the results displayed in Fig.~8, the input
interaction potential required a large diffuseness consistent  with  the target
having an extended neutron matter distribution.  The results on the far left in
Fig.~8, were found with the deformation set to zero.              They show the
sub-structural origins of each state in the actual spectrum~\cite{Pi05}.  Again
use of the OPP to account for Pauli blocking of occupied states  was crucial in
finding these results. With zero deformation then states reduce to those of   a
bound proton to one of the three states of ${}^6$He.      The ground state is a
$0p_{3/2}$ proton bound to the ${}^6$He ground and 1.8 MeV above  that there is
the  quadruplet of states formed by such a proton bound to the  first   excited
($2^+$) state.   That is also the case for the quadruplet  5.6  MeV  above  the 
ground and due to coupling to the second excited ($2^+$) state.      The single
state remaining is that of a $0p_{1/2}$ proton bound to the ground.        With
non-zero deformation, these states mix and spread to give the theoretical   set
that    compare  quite favorably with those known from experiment for ${}^7$Li. 
Evaluating  the  $n$+${}^6$He system, one would expect more sub-threshold states, 
as is the case with the mass 13 systems. \\
\begin{minipage}[t]{6.5cm}
\vspace*{1mm}
\hspace*{-0.5cm}

\epsfig{figure=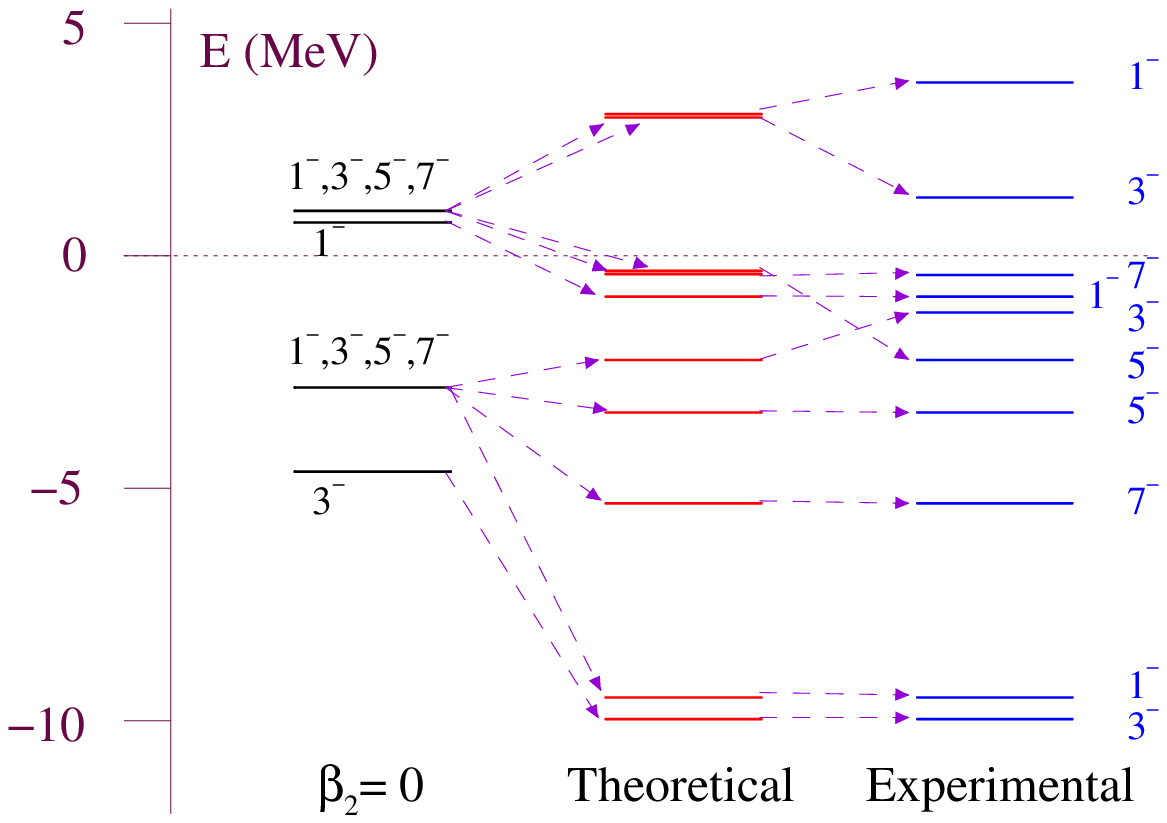,clip=,width=0.97\linewidth}

\vspace*{1mm}
\end{minipage} \hfill
\begin{minipage}[t]{6.0cm}

However,  one also has to alter the OPP since there is a (partial at least) new 
blocking of the $0p$-shell.  The MCAS result for the particle-unstable
nucleus ${}^7$He gives  one weakly-bound state.      A slight adjustment of the
interaction chosen would move that state into the continuum without   seriously
changing the ${}^7$Li results.\\

\vspace*{1mm}
{\small Figure~8:
The spectra of ${}^7$Li as determined using MCAS for the $p$+${}^6$He system.}
\vspace*{1mm}
\end{minipage} \hfill

\section{MCAS and the optical potential}

  Assuming a local form for the elastic channel element of the potential matrix 
in the MCAS approach, and with the operator 
\begin{equation}
{\mathbf \Lambda}(E) =
\left[ {\mathbf \eta} - {\bf G}_0^{(Q)}(E)\right]^{-1}
- {\mathbf \eta}^{-1}\  ,
\end{equation}
involving the free Green function excluding the elastic channel,   the  optical potential 
for elastic scattering takes the nonlocal form~\cite{Am03} with the addition of
\begin{equation}
\Delta U(r,r'; E) =
\sum^N_{n, n' = 1} \chi_{1 n}(r)
\left[{\mbox{\boldmath $\Lambda$}}(E) \right]_{nn'}(E) \chi_{1 n'}(r')\ ,
\end{equation}
to $V_{11}(r)$. Here
$\chi_{1 n}(r)$ are Bessel transforms of the form factors $\chi_{1 n}(k)$ and
$\Delta U$ is known as the dynamic polarization potential (DPP).        At low
energies it is this makes the formulated optical potential complex      (if the
energy allows more than one open channel), nonlocal, and energy dependent.  The
DPP has been calculated using the\\
\begin{minipage}[t]{5.0cm}
\vspace*{1mm}

\epsfig{figure=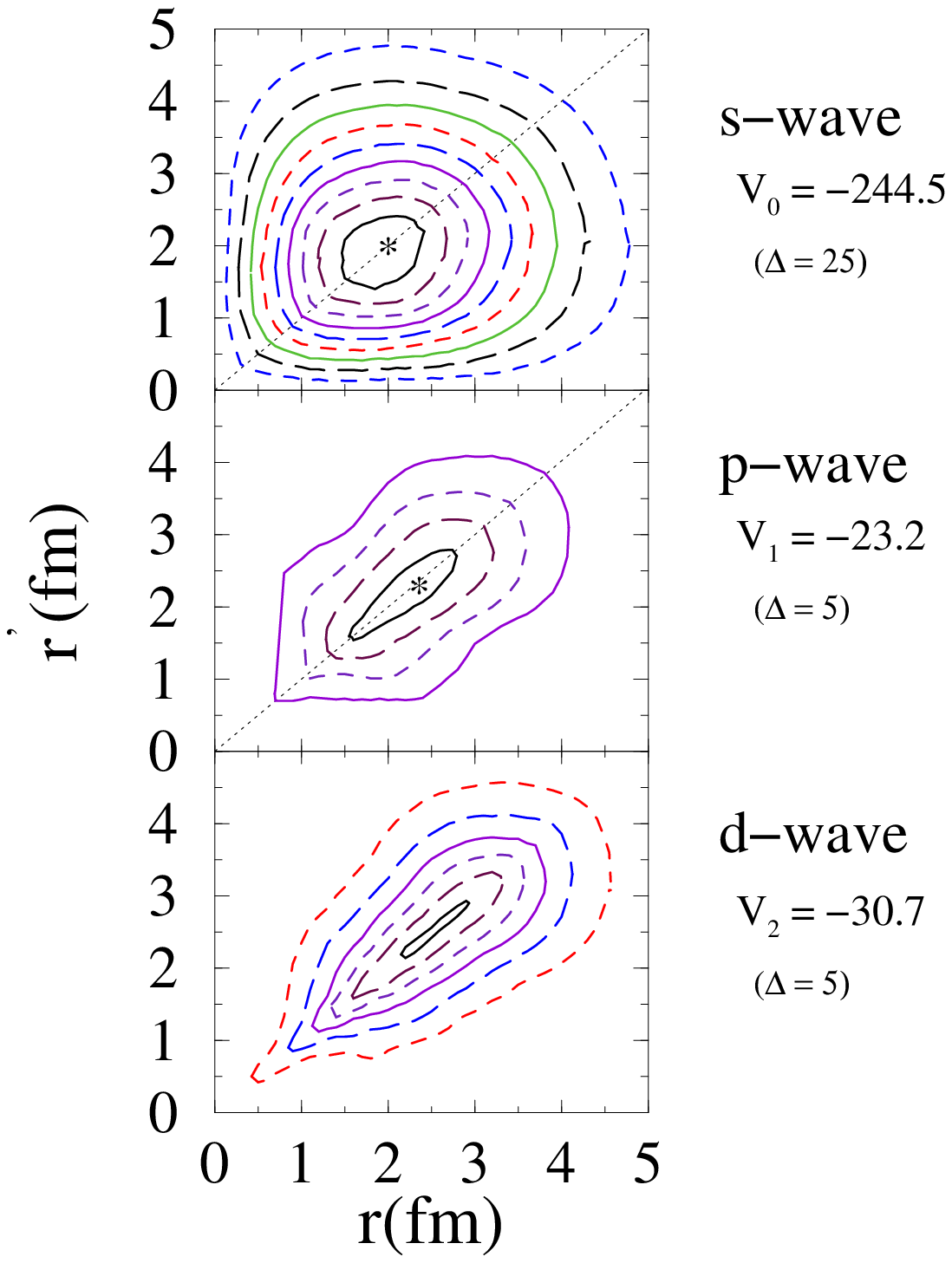,clip=,width=1.1\linewidth}

\end{minipage} \hfill
\begin{minipage}[t]{7.0cm}

MCAS approach and values of them  found for 2.73 MeV neutrons on ${}^{12}$C
are plotted in Fig.~9 for three partial waves, $\ell = 0, 1, 2$.    The central
depths  of  each  are  indicated  on  the  right as are the energy steps of the 
contour lines. As the energy is below the first inelastic threshold, the DPP is
purely real.    But the coupled-channel effects make that aspect of the optical 
potential extremely nonlocal. The DPP is also very angular momentum dependent.
Of course for energies above 4.4389 MeV,      the DPP becomes complex since the 
threshold for the first inelastic scattering  has been exceeded.\\

{\small Figure~9:  The  DPP  (MeV-fm$^{-1}$)  for three partial  waves  for 
2.73 MeV neutrons from ${}^{12}$C}
\end{minipage}\hfill

\section{Conclusions}

  The MCAS approach to analyze (low-energy) nucleon-nucleus scattering is built 
from a model structure of the interaction potentials between a nucleon and each
of the target states taken into consideration. All resonances (narrow and broad)
in the cross section within the selected (positive) energy range can  be  found
on a background.          With negative energies, the MCAS method specifies the
sub-threshold bound states of the compound nucleus.     The scheme allows for a
resonance finding procedure by which all resonances   (and sub-threshold bound)
states will be found within a chosen energy range.   Further via use of the OPP
in definition of the Sturmians to be used,   those functions are assured to be 
orthogonal to any single nucleon bound state that is Pauli blocked.        That
allowance   for   the   influence  of the Pauli principle is crucial in finding 
results that concur well with observation.


\end{document}